\documentclass[11pt,twoside,a4paper,cmspaper,final,collab]{cms-tdr}
\def\svnVersion{e333937-D}\def\svnDate{2026/07/27}\def\cmsCernNoTag{CERN-EP-2026-184}\def\cmsCernDate{\today}\def\cmsMessage{Submitted to Physics Letters B}
\begin{document}\cmsNoteHeader{SMP-24-013}

\newlength\cmsFigWidth
\newlength\cmsTabSkip\setlength{\cmsTabSkip}{1ex}

\newcommand{\tauhadpm}{\HepParticle{\PGt}{\mathrm{h}}{\pm}\xspace}
\newcommand{\tauhad}{\HepParticle{\PGt}{\mathrm{h}}\xspace}
\newcommand{\Pj}{\ensuremath{{\mathrm{j}}}\xspace}

\cmsNoteHeader{SMP-24-013}

\title{Combination of vector boson scattering measurements in leptonic final states in proton-proton collisions at  \texorpdfstring{$\sqrt{s} = 13\TeV$}{sqrt(s) = 13 TeV}}

\author[cern]{A. Cern Person}

\date{\today}

\abstract{
   A statistical combination of vector boson scattering measurements in proton-proton collisions at $\sqrt{s} = 13\TeV$ is presented. The analysis is based on a dataset corresponding to an integrated luminosity of 138\fbinv, collected with the CMS detector in 2016 -- 2018. Events are selected to contain two vector bosons, with at least one decaying leptonically, and at least two additional jets with a high invariant mass and a large pseudorapidity separation. Signal rates are extracted for same-sign $\PW\PW$, opposite-sign $\PW\PW$, as well as for $\PW\PZ$ and $\PZ\PZ$ channels, and are compared with the standard model predictions. The combination provides a global view of vector boson scattering by simultaneously constraining all channels in a coherent framework, improving sensitivity beyond individual inputs, with all results in agreement with the standard model predictions. In addition, a charge-separated measurement is performed in which $\PW$ bosons are distinguished according to their electric charge, allowing sensitivity to the production charge asymmetry.
}

\hypersetup{
pdfauthor={CMS Collaboration},
pdftitle={Combination of vector boson scattering measurements in leptonic final states in proton-proton collisions at  sqrt{s} = 13 TeV},
pdfsubject={CMS},
pdfkeywords={CMS, VBS, combination, SM}}

\maketitle 

\section{Introduction}

Since the discovery of the Higgs boson by the ATLAS and CMS Collaborations in 2012~\cite{ATLAS_Higgs,CMS_Higgs,CMS:2013btf}, its properties have been studied in depth~\cite{CMS_10_years, ATLAS_10_years} showing that the observations are in good agreement with the Brout--Englert--Higgs mechanism~\cite{Englert:1964et,Higgs:1964ia,Higgs:1964pj,Guralnik:1964eu,Higgs:1966ev,Kibble:1967sv}. 
While this discovery is consistent with the standard model (SM) description of spontaneous electroweak symmetry breaking (EWSB), our understanding of the Yukawa and scalar sectors remains incomplete. Studying the high-energy behavior of electroweak (EW) $\PW$ and $\PZ$ bosons through vector boson scattering (VBS) offers important insights into the EWSB mechanism. 
At energies far above the EW scale, the longitudinal modes of weak bosons represent the Nambu--Goldstone bosons resulting from the spontaneous EWSB~\cite{Veltman:1989ud,He:1992nga}. When considering only gauge interactions, the scattering amplitude for longitudinally polarized weak bosons increases with energy, leading to a violation of unitarity at center-of-mass energies of about 1\TeV~\cite{PhysRevD.16.1519,PhysRevLett.38.883}. 
The inclusion of scalar contributions to the scattering amplitudes, mediated by the SM Higgs boson, restores unitarity at high energies. This delicate cancellation could be altered by phenomena beyond the SM, which may modify the coupling strength of the Higgs boson to gauge bosons~\cite{Espriu:2012ih,Chang:2013aya} or alter the triple and quartic self-interactions of gauge bosons~\cite{Eboli:2006wa}, stemming from the non-Abelian gauge structure of the EW sector. The new physics contributions could manifest in VBS as an excess of events compared to the SM predictions.

In proton-proton ($\Pp\Pp$) collisions at the CERN LHC, the VBS process occurs when the initial-state partons each radiate a vector boson, which then scatter off each other. The initial-state partons are scattered at small angles, resulting in highly energetic jets produced in the forward and backward directions. 
At tree level in perturbation theory, the VBS signature arises from purely EW interactions at order $\alpha_{\text{EW}}^4$, where $\alpha_{\text{EW}}$ counts the order of EW vertices in the amplitude, including both charged and neutral currents. A topologically similar background arises from processes induced by quantum chromodynamics (QCD) at order $\alpS^2 \alpha_{\text{EW}}^2$, where \alpS is the strong coupling.
Given the weakness of $\alpha_{\text{EW}}$, EW-induced VBS production is one of the rarest processes studied at the LHC.

The scattering of massive vector bosons ($\PW$, $\PZ$) was first observed in the $\PWpm\PWpm$ state~\cite{CMS:2017fhs} and the observation of all states was completed with the measurements of the EW production of pairs of $\PZ$ bosons by the ATLAS~\cite{ATLAS:2020nlt} and CMS collaborations~\cite{CMS:2026zxi}.

This Letter presents the most comprehensive statistical combination of VBS measurements using $\Pp\Pp$ collisions at a center-of-mass energy of $\sqrt{s} = 13\TeV$ recorded by the CMS experiment. For all considered final states, the dataset corresponds to an integrated luminosity of 138\fbinv collected in 2016 -- 2018~\cite{CMS:2021xjt,CMS:LUM-17-004,CMS:LUM-18-002}. The measurements focus on the EW production of massive vector bosons in final states where these bosons decay into four fermions ($\PZ\to\Pellp\Pellm, \PQq\PAQq$ and $\PW\to\Pellpm\PGn,\tauhadpm\PGnGt,\PQq\PAQq$, where $\Pell = \Pe$ or $\PGm$ and where $\tauhadpm$ is a tau lepton that decays hadronically), accompanied by the scattered partons.

Two combination models are defined to measure the EW production of two heavy vector bosons plus two jets in the VBS topology. The first model targets the simultaneous measurement of four signal strengths gauging the production cross section of two $\PW$ bosons with same ($\PWpm\PWpm$) or opposite ($\PWp\PWm$) electric charge, $\PWpm\PZ$, and $\PZ\PZ$. The signal strength are defined as the ratio of the observed rates in the various final states to the SM predictions.
The second model differentiates the $\PW$ boson electric charge, presenting six parameters to be measured simultaneously: $\PWp\PWp$, $\PWm\PWm$, $\PWp\PWm$, $\PWp\PZ$, $\PWm\PZ$, and $\PZ\PZ$. While the former provides a global investigation of VBS processes, with direct comparison to existing results, the latter extends the analysis by explicitly probing the expected production charge asymmetry. In $\Pp\Pp$ collisions, the differing parton distribution functions (PDFs) of up and down quarks (stemming from the valence quark composition of the proton) lead to an inherent asymmetry in the production rates, favoring positively charged final states~\cite{ATLAS:2016nqi,ATLAS:2019fgb,CMS:2012ivw,CMS:2013pzl,CMS:2016qqr,LHCb:2012lki,LHCb:2015mad,ATLAS:2025hhn,ATLAS:2025ede,CMS:2020cph}.
 
\begin{figure*}[!htb]
   \centering
   \includegraphics[width=\textwidth]{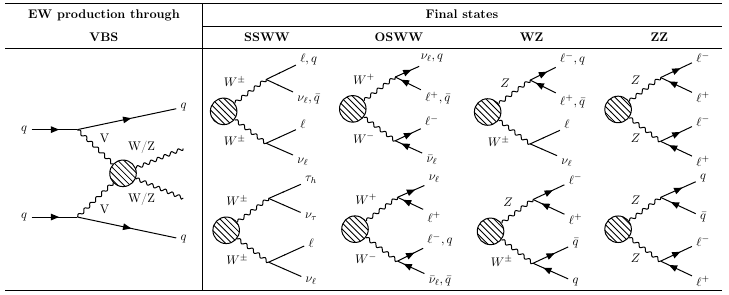}
   \caption{The left side shows a representative schematic diagram for the EW-induced production of VBS involving massive vector bosons. The dashed circle encapsulates different contributions leading to the same final state, including triple and quartic gauge couplings, as well as Higgs boson exchange. On the right, the analyzed decay channels are grouped into same-sign and  opposite-sign $\PW\PW$, and $\PW\PZ$, $\PZ\PZ$ final states, including both leptonic and hadronic decays of the heavy vector bosons. In all cases, at least one vector boson is required to decay leptonically. Lepton lines in the $\PWpm$ decay vertices carry no arrows or charge labels, as a single diagram represents both the $\PWp$ and $\PWm$ decay modes. }\label{fig:feyn_diags}
 \end{figure*}

By denoting $\Pell = \Pe,\PGm$, the fully leptonic final states included in the combination are the production of two $\PW$ bosons with same electric charge $\PWpm\PWpm \to \Pellpm\Pellpm{}2\PGn$~\cite{CMS:2020gfh} (SS$\PW\PW$) or with a hadronically decaying $\PGt$ lepton $\PWpm\PWpm \to \Pellpm\tauhadpm{}2\PGn$~\cite{CMS:2024jvu}, two $\PW$ bosons with opposite electric charge $\PWp\PWm \to \Pellp\Pellm{}2\Pgn$~\cite{CMS:2022woe} (OS$\PW\PW$), associated $\PWpm\PZ \to 3\Pell\PGn$ production~\cite{CMS:2020gfh} ($\PW\PZ$), and $\PZ\PZ\to 4\Pell$~\cite{CMS:2020fqz} ($\PZ\PZ$). 
Additionally, two processes in which one heavy vector boson decays hadronically, $\PW\PV\to \Pell\PGn\Pj\Pj$~\cite{CMS:2021qzz} and $\PZ\PV\to 2\Pell \Pj\Pj$~\cite{CMS:2025dbm}, are included. Here, $\PV=\PW,\PZ$ denotes the hadronically decaying vector boson and $\Pj$ the quark jet. These channels enhance the overall sensitivity by providing complementary kinematic information to the fully leptonic final states. Due to the limited invariant mass resolution of the hadronic decay of the vector boson ($\PV\to\Pj\Pj$), these channels do not allow for a clear separation of individual VBS production modes. As a result, they probe a combination of VBS processes and contribute to the measurement of multiple cross section parameters simultaneously. Figure~\ref{fig:feyn_diags} shows the schematic diagram for the EW VBS production of massive vector bosons, as well as the decay channels of interest in this Letter.

The EW production of a massive vector boson in association with a photon~\cite{CMS:2022yrl,CMS:2021gme} probes a topology distinct from the production of two massive vector bosons. As a result, these measurements are not included in the combined interpretation presented here.

Tabulated results are provided in the HEPData record for this analysis~\cite{hepdata}.

\section{The CMS detector}

A key component of the CMS detector is a superconducting solenoid with a 6\unit{m} internal diameter, producing a magnetic field of 3.8\unit{T}. 
Inside the solenoid, the detector houses a silicon pixel and strip tracker~\cite{CMS:2014pgm,CMS:2021ime}, a lead-tungstate crystal electromagnetic calorimeter (ECAL), and a brass and scintillator hadron calorimeter (HCAL), each consisting of a barrel and two endcap sections. Forward calorimeters extend the pseudorapidity ($\eta$) coverage of the barrel and endcaps up to $\abs{\eta} < 5$. Muons are detected using gas-ionization chambers embedded within the steel flux-return yoke outside the solenoid. 
The particle-flow algorithm~\cite{CMS:2017yfk} is employed to reconstruct and identify individual particles in each event, optimizing the combination of information from the various components of the CMS detector. 
The electron energy is determined from a combination of the electron's momentum at the primary interaction vertex (as measured by the tracker), the energy of the associated ECAL cluster and the energy sum of all bremsstrahlung photons spatially linked to the electron track~\cite{CMS:2020uim}. The muon four-momentum is derived from the curvature of their corresponding track as measured in the tracker and in the muon detectors~\cite{CMS:2018rym}. Charged-hadron energy is determined by combining the momentum measured by the tracker with the matching ECAL and HCAL energy deposits, with corrections for the calorimeters response to hadronic showers. Finally, neutral hadron energy is obtained from the corresponding corrected ECAL and HCAL measurements.
Events of interest are selected by a two-tiered trigger system. The first level, based on custom hardware processors,
uses data from the calorimeters and muon detectors to select events at a rate of approximately 100\unit{kHz}, with a latency of 4\mus~\cite{CMS:2020cmk}. The second level, known as the high-level trigger, 
is a processor farm running a fast version of the full event reconstruction software, reducing the event rate to a few \unit{kHz} before data storage~\cite{CMS:2016ngn,CMS:2020cmk,CMS:2024aqx}. A more comprehensive description of the CMS detector, along with details of the coordinate  system and relevant kinematic variables, can be found in Refs.~\cite{CMS:2008xjf,CMS:2023gfb}.

\section{Input analyses}

The target production modes and final states considered are listed in Table~\ref{tab:analyses}. More details on each of the input analyses can be found in the references listed in the last column. A summary of the event selection, categorization, and phase space orthogonality is given in Section~\ref{sec:event_selection}.

\begin{table}[!htp]
   \centering
      \topcaption{Summary of the analyses considered in the combination. The symbol $\Pell$ denotes either an electron or muon, $\tauhad$ a hadronically decaying $\PGt$ lepton, and $\Pj$ is the shorthand notation for a jet.\label{tab:analyses} }
      \renewcommand{\arraystretch}{1.2}
      \begin{tabular}{ l c c } 
      Shorthand name &  Final state & Reference \\
      
      \hline
      SS$\PW\PW$\,($\Pe$, $\PGm$)  & $\Pellpm\Pellpm{}2\Pgn \Pj\Pj$   & \cite{CMS:2020gfh} \\
      SS$\PW\PW$\,(\tauhad)     &   $\Pellpm\tauhadpm{}2\PGn \Pj\Pj$  & \cite{CMS:2024jvu} \\
      OS$\PW\PW$                & $\Pellpm\Pellmp{}2\Pgn \Pj\Pj$   & \cite{CMS:2022woe} \\
      $\PW\PV$                  & $\Pell\PGn\Pj\Pj\Pj\Pj$  & \cite{CMS:2021qzz} \\
      $\PW\PZ$                  & $3\Pell\PGn\Pj\Pj$   & \cite{CMS:2020gfh} \\
      $\PZ\PV$                  & $2\Pell\Pj\Pj\Pj\Pj$ & \cite{CMS:2025dbm}\\
      $\PZ\PZ$\,($4\Pell$)         & $4\Pell\Pj\Pj $ &\cite{CMS:2020fqz} \\
      \end{tabular}
\end{table}

The processes of interest are sensitive to a combination of EW production modes involving two heavy vector bosons and two jets in the VBS topology. The analyses can be categorized based on the targeted VBS production mode and the decay channels of the vector bosons. Analyses focusing on the fully leptonic decays of the vector boson pair ($\PW \to \Pell\PGn$, $\PZ \to \Pell\Pell$) are sensitive to a single VBS production mode and dominate the sensitivity, thanks to their clean signature inside the detector. In SS$\PW\PW$, sensitivity is enhanced by combining two decay channels with different experimental signatures: one where both $\PW$ bosons decay into electrons or muons, referred to as SS$\PW\PW$\,($\Pe$, $\Pgm$), and one where a $\PW$ boson decays into a hadronically decaying $\PGt$ lepton, denoted as SS$\PW\PW$\,($\tauhad$). All channels leverage hadronic decays of the heavy vector bosons to further constrain the production rates. Specifically, the $\PW\PV$ final state, where $\PW \to \Pell\PGn$ and $\PV(\PW,\PZ) \to \Pj\Pj$, captures a mixture of production modes, including SS$\PW\PW$ scattering, OS$\PW\PW$, and the scattering of a $\PW\PZ$ pair. Another channel, referred to as $\PZ\PV$, targets $\PZ \to \Pell\Pell$ and $\PV(\PW,\PZ) \to \Pj\Pj$, and complements the $\PW\PV$ one by providing additional constraining power for the EW production of $\PW\PZ$ and $\PZ\PZ$ pairs. 

\section{Signal and background samples}\label{sec:sig_bkg_simulation}

The EW-induced VBS signals are generated at leading order (LO) with \MGvATNLO\,~\cite{Alwall:2014hca,Alwall:2007fs} (versions 2.4.6 and 2.6.5). The decays of the vector bosons are simulated with \textsc{MadSpin}~\cite{Madspin}, accounting for finite-width effects and spin correlations.

The estimation of the background contributions varies across final states and combines both Monte Carlo (MC)  simulation and techniques based on control samples in data. Reducible or poorly simulated backgrounds, such as nonprompt leptons, charge-misidentified electrons, and $\PW$/$\PZ$+jets processes, are constrained using dedicated control regions (CRs) in data.

Most irreducible backgrounds are modeled using MC simulation. Predictions for the QCD-induced background and EW-QCD interference are simulated in the same way as the VBS signals, with the exception of the OS$\PW\PW$ quark- and gluon-induced $\PW\PW$ background, which is simulated using \POWHEG v2.0~\cite{Nason:2004rx,Frixione:2002ik,Frixione:2007vw,Alioli:2008gx,Alioli:2010xd}. The EW-QCD interference is included as part of the EW signal only in SS$\PW\PW$\,($\Pe$, $\PGm$) and $\PZ\PZ$ analyses, while it is negligible otherwise.

Other processes are simulated as follows: $\PZ$/$\PW$+jets, $\PW\Pgg$, QCD-induced diboson, and vector boson fusion production at LO with \MGvATNLO as well as triboson production at next-to-LO (NLO) and $\PZ\Pgg$ at LO or NLO depending on analysis; top quark processes at NLO mainly with \POWHEG or \MGvATNLO; diboson at LO with \MCFM~\cite{Campbell:2015qma,Campbell:2011bn,Campbell:1999ah,Campbell:2019dru} or \MGvATNLO; diboson at NLO with \POWHEG or \MGvATNLO. Higgs boson production, treated as a background, uses \POWHEG or \textsc{JHUGen}~\cite{Bolognesi:2012mm} at NLO with reweighting to next-to-NLO (NNLO)~\cite{Hamilton:2015nsa}, where applicable.

The PDFs are taken from the NNPDF sets~\cite{NNPDF:2014otw, NNPDF:2017mvq}, using version 3.0 or 3.1 at NLO and NNLO, respectively, depending on the analysis and the data-taking year.

Parton showering, hadronization, and the underlying event are modeled with \PYTHIA~\cite{Sjostrand:2014zea,Bierlich:2022pfr} (versions 8.226, 8.230, or 8.306 depending on the analysis and data-taking year). Additional proton-proton interactions in the same or nearby bunch crossings (pileup) are simulated with \PYTHIA and are included in all MC sample. Dipole recoil~\cite{Cabouat:2017rzi} is applied in $\PW\PV$ and $\PZ\PV$ channels to improve the modeling of additional jets. The underlying event is simulated using tuned parameter sets, specifically the CUETP8M1~\cite{Skands:2014pea,CMS:2015wcf} or CP5~\cite{CMS:2019csb,CMS:2022awf} tune, depending on the analysis and the data-taking year.

All events are processed with \GEANTfour~\cite{GEANT4:2002zbu}, which simulates the passage of particles through the CMS detector and their interactions with the detector material. The simulated signals are then reconstructed using the standard CMS reconstruction algorithms~\cite{CMS:2017yfk}.

\section{Event selection and signal extraction}\label{sec:event_selection}

Single-electron and -muon triggers, requiring an isolated lepton, are employed to record collision events across all analyses included in the combination. The lepton transverse momentum ($\pt$) thresholds range from 25 to 35\GeV for electrons and 24 to 27\GeV for muons, depending on the final state, lepton pseudorapidity, and data-taking period. Except for the $\PW\PV$ and SS$\PW\PW$\,(\tauhad) analyses, an additional set of dilepton triggers with lower $\pt$ thresholds and looser isolation is used. In the $\PZ\PZ$\,($4\Pell$) analysis, these dilepton triggers serve as the primary selection, supplemented by three-lepton triggers to enhance efficiency. This combination achieves a trigger efficiency exceeding 97\% for all analyses.

Offline selections are applied to isolate the VBS topology and reduce background contributions, creating signal-enriched regions (SRs). Background-enriched CRs are defined by inverting specific selection criteria to improve the modeling of key backgrounds.

Mutually exclusive criteria are applied across all analyses to avoid counting events multiple times. The primary and most important mechanism used to enforce orthogonality between regions is the definition of exclusive charged-lepton multiplicities: each region is first defined by selecting events with a specific number of charged leptons ($\Pe$, $\Pgm$), and then explicitly vetoing events with additional charged leptons beyond that requirement. This procedure establishes a strict partition of the event sample into non-overlapping categories.

The VBS-enriched SRs require two jets ($\pt^{\Pj}>30$ or 50\GeV depending on the final state and $\abs{\eta_\Pj}<4.7$) with high invariant mass ($m_{\Pj\Pj}^{\text{VBS}} > 100$ or 500\GeV) and large pseudorapidity separation ($\abs{\Delta\eta^{\text{VBS}}_{\Pj\Pj}}>2.5$), along with at least one leptonically decaying vector boson. Additional selections tailored to each final state enhance the signal-to-background ratio.

After the orthogonality from lepton multiplicity and lepton vetoes is established, further exclusivity between regions is ensured through additional selections on kinematical observables and object properties. These include requirements on hadronically decaying tau leptons, total lepton charge, and same-flavor or different-flavor lepton pairs. Orthogonality is further refined using invariant mass windows for dilepton or vector boson candidates, and by separating resolved and boosted topologies with requirements on small- and large-radius jets, as well as on the reconstructed mass of the hadronically decaying vector boson. Requirements on jets originating from \PQb~quarks are also applied where relevant.

Together, these selections ensure that each region targets a specific signal or background process without overlap, with the charged-lepton multiplicity and veto strategy providing the fundamental, first layer of orthogonality, and kinematic selections providing additional refinement.

Due to the complex VBS topology and the large background contamination in the SRs, advanced discriminants are employed to enhance sensitivity to the EW VBS signal. The $\PW\PV$ and $\PZ\PV$ analyses, the OS$\PW\PW$ different-flavor category, and the SS$\PW\PW$\,(\tauhad) channel use classifiers based on deep neural networks (DNN), with binned output distributions serving as input for the statistical analysis. The $\PW\PZ$ analysis relies on a boosted decision tree (BDT), whereas the $\PZ\PZ$\,($4\Pell$) analysis uses a matrix element based discriminant~\cite{Bolognesi:2012mm,Gao:2010qx,Anderson:2013afp,Gritsan:2016hjl} ($K_D$) to separate EW- and QCD-induced contributions. Analyses with higher signal purity, such as SS$\PW\PW$\,($\Pe$, $\PGm$), exploit single- or double-differential kinematic observables derived from VBS-sensitive variables.

Table~\ref{tab:regions_observables} summarizes, for each analysis and region, the observable used in the combined fit and the number of bins in the input distribution.

\begin{table*}[!htb]
   \centering
       \topcaption{ Summary of the phase space regions entering the combined fit, with the corresponding observables and number of bins. Two-dimensional observables are indicated by a colon separating their two components. Unless the data-taking year is explicitly specified in the subregion column, separate templates are constructed for each year of data taking. DY denotes Drell-Yan processes. The $\PQb$-tag ($\PQb$-veto) category requires at least one (zero) $\PQb$-tagged jets~\cite{Bols:2020bkb} in the event. For WV and ZV, the resolved (merged) region targets events with four small-radius jets (one large-radius jet accompanied by two small-radius jets). DNN classifiers are labeled by the corresponding region name only when distinct models are used across analysis regions.
       \label{tab:regions_observables} }
       {
       \renewcommand{\arraystretch}{1.2}
       \begin{tabular}{ c  c  c  c  c  } 
      \hline
      Analysis & Region & Subregions & Observable & Bins \\ 
      \hline 

      SS$\PW\PW$\,($\Pe$, $\PGm$) & SR & - & $m_{\Pj\Pj}^{\text{VBS}}:m_{\Pell\Pell}$ & 32 \\
      SS$\PW\PW$\,($\Pe$, $\PGm$) & CR $\PQb$-tag & - & $m_{\Pj\Pj}^{\text{VBS}}$ & 4 \\

      SS$\PW\PW$\,(\tauhad) & SR & ($\Pe, \Pgm$) & DNN & 12 \\
      SS$\PW\PW$\,(\tauhad) & CR $\ttbar$ & ($\Pe, \Pgm$) & DNN & 12 \\
      SS$\PW\PW$\,(\tauhad) & CR OS & ($\Pe, \Pgm$) & DNN & 12 \\

      OS$\PW\PW$ & SR & $\Pe\Pgm$ ($\PZ_{\Pell\Pell} \lesseqgtr 1$ ) & DNN & 13 \\ 
      OS$\PW\PW$ & SR & $\Pe\Pe, \, \Pgm\Pgm$ ($\PZ_{\Pell\Pell} \lesseqgtr 1$ ) & $m_{\Pj\Pj}^{\text{VBS}}:\Delta\eta_{\Pj\Pj}^{\text{VBS}}$ & 8 \\ 
      OS$\PW\PW$ & CR $\ttbar$ & ($\Pe\Pe$, $\Pgm\Pgm$, $\Pe\Pgm$) & Events & 1 \\
      OS$\PW\PW$ & CR DY & ($\Pe\Pgm$) & Events & 1 \\
      OS$\PW\PW$ & CR DY & ($\Pe\Pe$, $\Pgm\Pgm$, $\abs{\Delta\eta_{\Pj\Pj}^{\text{VBS}}}\lesseqgtr 5$) & Events & 1 \\

      $\PW\PV$ & SR resolved & ($\Pe,\, \Pgm$) & DNN resolved & 25 \\
      $\PW\PV$ & SR merged & ($\Pe,\, \Pgm$) & DNN boosted & 12 \\
      $\PW\PV$ & CR $\ttbar$ resolved & ($\Pe,\, \Pgm$) & Events & 1 \\
      $\PW\PV$ & CR $\ttbar$ merged & ($\Pe,\, \Pgm$) & Events & 1 \\
      $\PW\PV$ & CR $\PW$+jets resolved & ($\Pe,\, \Pgm$) & $\pt^{\PW\to \Pell\Pgn}$ & 21 \\
      $\PW\PV$ & CR $\PW$+jets merged & ($\Pe,\, \Pgm$) & $\pt^{\PW \to \Pell\Pgn}:\pt^{\Pj2, \text{VBS}}$ & 7 \\

      $\PW\PZ$ & SR & - & BDT & 8 \\
      $\PW\PZ$ & CR $\PQb$-tag  & - & $m_{\Pj\Pj}^{\text{VBS}}$ & 4 \\

      $\PZ\PV$ & SR resolved $\PQb$-tag & - & DNN resolved $\PQb$-tag & 21 \\
      $\PZ\PV$ & SR merged $\PQb$-tag & -  & DNN boosted $\PQb$-tag & 21 \\
      $\PZ\PV$ & SR resolved $\PQb$-veto & - & DNN resolved $\PQb$-veto & 21 \\
      $\PZ\PV$ & SR merged $\PQb$-veto & - & DNN boosted $\PQb$-veto & 21 \\

      $\PZ\PV$ & CR DY resolved  & $\PQb$-tag, $\PQb$-veto (2017-2018) & $\pt^{\PZ \to \Pell\Pell}:\pt^{\Pj2, \text{VBS}}$ & 12\\
      $\PZ\PV$ & CR DY resolved &  $\PQb$-tag, $\PQb$-veto (2016) & $\pt^{\PZ \to \Pell\Pell}$ & 5 \\
      $\PZ\PV$ & CR DY merged  & $\PQb$-tag, $\PQb$-veto & $\pt^{\PZ \to \Pell\Pell}$ & 5 \\

      $\PZ\PZ$\,($4\Pell$) & SR & ($4\Pe, 4\Pgm, 2\Pe2\Pgm$) & $K_D$ discriminant & 25 \\

      \hline
      \end{tabular}
       }
\end{table*}

\section{Systematic uncertainties}

Systematic uncertainties from experimental and theoretical sources are derived and taken into account in the combined fit.

The integrated luminosity of the 2016, 2017, and 2018 datasets is known with uncertainties of 1.2, 2.3, and 2.5\%, respectively, resulting in an overall uncertainty of 1.6\%~\cite{CMS:2021xjt,CMS:2021xjt,CMS:LUM-17-004,CMS:LUM-18-002}. Lepton reconstruction and identification efficiencies, along with electron and muon momentum scale and resolution uncertainties, are evaluated using dedicated studies~\cite{CMS:2017wtu}. Trigger efficiency uncertainties are derived from measurements based on control samples in data and propagated to the predicted yields.

Jet energy scale and resolution uncertainties are assessed by varying the jet momenta within their uncertainties, affecting jet multiplicities and $\ptmiss$~\cite{CMS:2016lmd}. For small- and large-radius jets, additional uncertainties in the jet mass scale and resolution are considered, particularly in channels targeting boosted topologies. 

Discrepancies in jet pileup identification between data and simulation are corrected by multiplicative factors applied to all simulated samples~\cite{CMS:2020ebo}. These corrections affect only analyses with jets of $\pt < 50$ GeV. The uncertainties on these correction factors are propagated to the fit and parameterized with one nuisance parameter per year, correlated across processes. Efficiencies for \PQb~tagging are adjusted using corrections based on control samples in data~\cite{CMS:2017wtu}. The uncertainties on these corrections are propagated to the fit and parameterized as nuisance parameters, correlated across processes but uncorrelated across data-taking years. Pileup modeling uncertainties are estimated by varying the total inelastic $\Pp\Pp$ cross section by 4.6\%~\cite{CMS:2018mlc}. Nonprompt-lepton background uncertainties account for differences between measurement and SRs, with separate components addressing systematic effects and statistical fluctuations from limited sample sizes.

Theoretical uncertainties are propagated by varying the renormalization and factorization scales up and down by a factor of two, excluding the two extreme and opposite variations, and by evaluating uncertainties from PDFs and \alpS following the PDF4LHC prescription~\cite{Butterworth:2015oua}. Parton shower modeling uncertainties are derived by varying initial- and final-state radiation weights in the simulated samples.

Theoretical uncertainties are treated as correlated across analyses and between signal and background samples wherever appropriate. Experimental uncertainties are also correlated across analyses when justified. The adopted correlation scheme has been tested for robustness, with the results showing only limited sensitivity to variations in the correlation assumptions.

\section{Statistical procedure}\label{sec:statistics}

The statistical interpretation of the data follows the methodology shared by the ATLAS and CMS Collaborations~\cite{ATLAS:2011tau,ATLAS:2016neq}. The results were derived using the CMS statistical analysis tool, \textsc{Combine}~\cite{CMS:2024onh}. A profile likelihood ratio $q$ is used as a test statistic in order to measure the parameters of interest (POIs) and their associated confidence intervals. In this study, the effect of the POIs is to scale the normalization of the signal processes of interest. This scaling is represented by $\vec{\mu}$, whose entries are the per-channel signal strengths $\mu = \sigma_{\text{obs}}/\sigma_{\text{SM}}$; two parameterizations with 4 and 6 POIs are used, as detailed below. Systematic uncertainties affecting the extraction of the POIs are implemented as nuisance parameters $\vec{\theta}$ in the likelihood and can modify either the normalization or the shape of the predicted distributions for signal and background processes. Statistical uncertainties due to the finite size of simulated samples are treated using the Barlow--Beeston method~\cite{Barlow:1993dm,Conway:2011in}, with independent nuisance parameters for each bin. By denoting with $\vec{x}$ the measurements, the test statistic is defined as:
\begin{equation}\label{eq:profile_likelihood}
   q(\vec{\mu}) = -2\ln \frac{L(\vec{x} \, | \,  \vec{\mu},\hat{\hat{\vec{\theta}}})}{L(\vec{x} \, | \,  \hat{\vec{\mu}},\hat{\vec{\theta}})},
\end{equation}
where $L$ represents the likelihood function, and $\hat{\vec{\mu}}$ and $\hat{\vec{\theta}}$ in the denominator are the maximum likelihood estimators for the POIs and nuisance parameters, respectively. The term $\hat{\hat{\vec{\theta}}}$ in the numerator denotes the value of $\vec{\theta}$ that maximizes $L$ for a given $\vec{\mu}$.
The binned likelihood function $L$ is constructed as the product of Poisson probabilities for each bin of the input templates. It incorporates systematic uncertainties and their correlations, which are described by likelihood functions ($\pi$) associated with the estimation of the corresponding parameters, denoted $\vec{\widetilde{\theta}}$. The formulation of the likelihood function for a specific phase space is detailed in Eq.~(\ref{eq:likelihood}). 
Given the statistical independence of the data points $\vec{x}$ across different phase spaces included in the combination, the complete likelihood is obtained as the product of the individual per-channel likelihoods,
\begin{equation}\label{eq:likelihood}
   L(\vec{x} \, | \,\vec{\mu},\vec{\theta}) = \prod_{\text{bin}=k} \frac{(\vec{\mu}\cdot \vec{s}_k + b_k)^{n_k}}{n_k !} \re^{-(\vec{\mu}\cdot \vec{s}_k + b_k)} \, \prod_{\text{syst}=j} \pi(\widetilde{\theta}_j \, | \,\theta_j)
\end{equation}

\begin{figure*}[!htb]
   \centering
   \includegraphics[width=\textwidth]{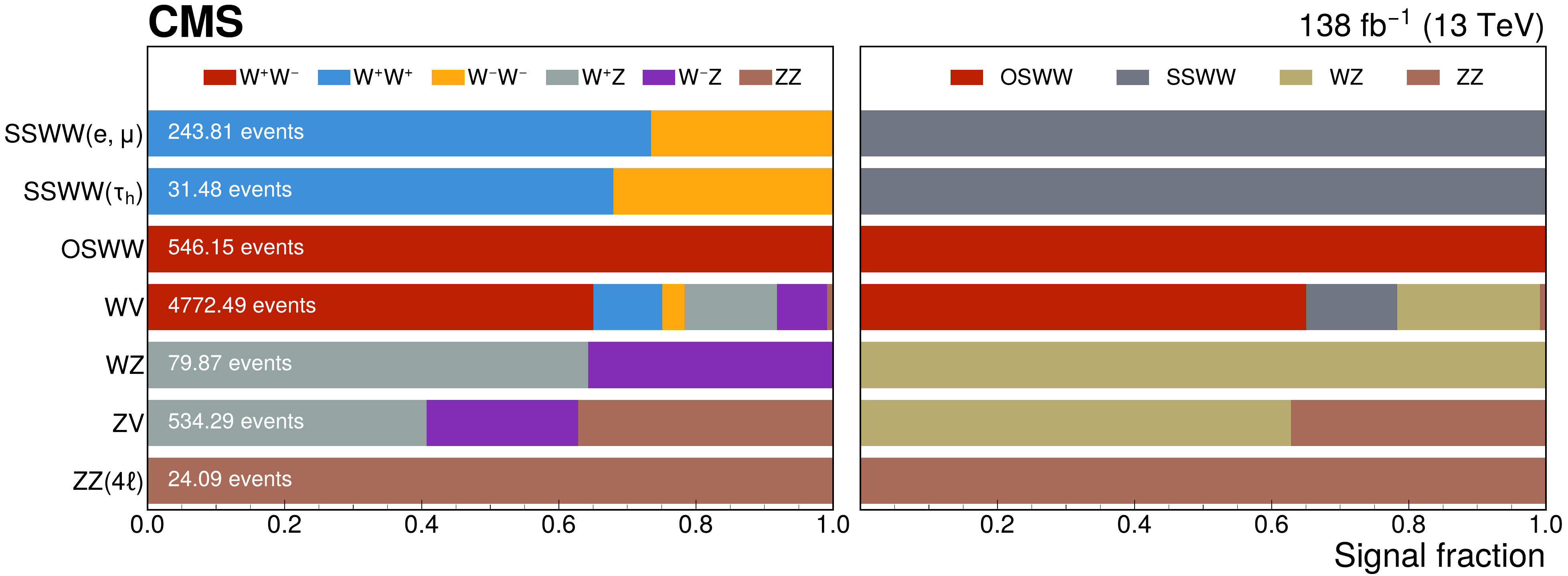}
   \caption{The signal fractions for each analysis contributing to the combination are presented, as well as the total expected signal yield for each channel. Rows correspond to the names of the analyses, while the $x$-axis indicates the signal fraction. Different colors, as shown in the legend, represent the distinct signal parameters targeted for measurement. The left half of the figure displays the signal fractions for the six components, separated by the electric charge of the W bosons. The right half of the figure provides the equivalent information for the four components. All signal yields are obtained from MC simulations prior to the combined fit of the signal strengths and include all analysis regions. The uncertainties on these yields are not shown in the figure.}\label{fig:sig_fractions}
 \end{figure*}

Confidence intervals corresponding to one and two standard deviations, equivalent to 68.3 and 95.4\% confidence levels (\CL{}s), are identified using the test statistic $q$. For a single POI, the intervals correspond to $q < 1$ and $q < 4$, respectively. For two POIs, these intervals are defined by $q < 2.3$ and $q < 6.18$, respectively. The statistical significance of a signal is computed using the profile likelihood ratio of Eq.~(\ref{eq:profile_likelihood}) where the null hypothesis corresponds to $\mu = 0$ as described in Refs.~\cite{CMS:2024onh,ATLAS:2011tau}. The distribution of the test statistic is determined using Wilk's theorem in the asymptotic limit~\cite{Cowan:2010js} and the observed $p$-value~\cite{Gross:2010qma} is converted into a statistical significance ($\sigma$) using a standard normal distribution~\cite{ParticleDataGroup:2024cfk}.

Two combination models are constructed: one with four parameters of interest (4-POIs) and another with six (6-POIs). The 4-POIs model includes four signal strengths that scale the normalizations of the EW VBS production processes: $\vec{\mu} = \left( \mu_{\mathrm{SS}\PW\PW},\mu_{\mathrm{OS}\PW\PW},\mu_{\PW\PZ},\mu_{\PZ\PZ} \right) $. The 6-POIs model, used to probe the production charge asymmetry, extends this parameterization by distinguishing the electric charge of the \PW bosons, introducing six signal strengths: $\vec{\mu} = \left( \mu_{\PWp\PWp},\mu_{\PWm\PWm},\mu_{\PWp\PWm},\mu_{\PWp\PZ},\mu_{\PWm\PZ},\mu_{\PZ\PZ}\right)$. Figure~\ref{fig:sig_fractions} illustrates the expected signal fractions for each of the input analyses, before the combined fit, separately for each of the two statistical models.

The CRs are included in the fit to constrain background normalizations and propagate corrections based on control samples in data to the SRs. The final combination to measure the VBS parameters of interest uses 48 phase space regions, with one template per region for each of the three data-taking years. Across all years, the total number of bins entering the combined fit is 1525.

\section{Results}

\begin{figure*}[!htb]
    \centering
    \includegraphics[width=0.45\textwidth]{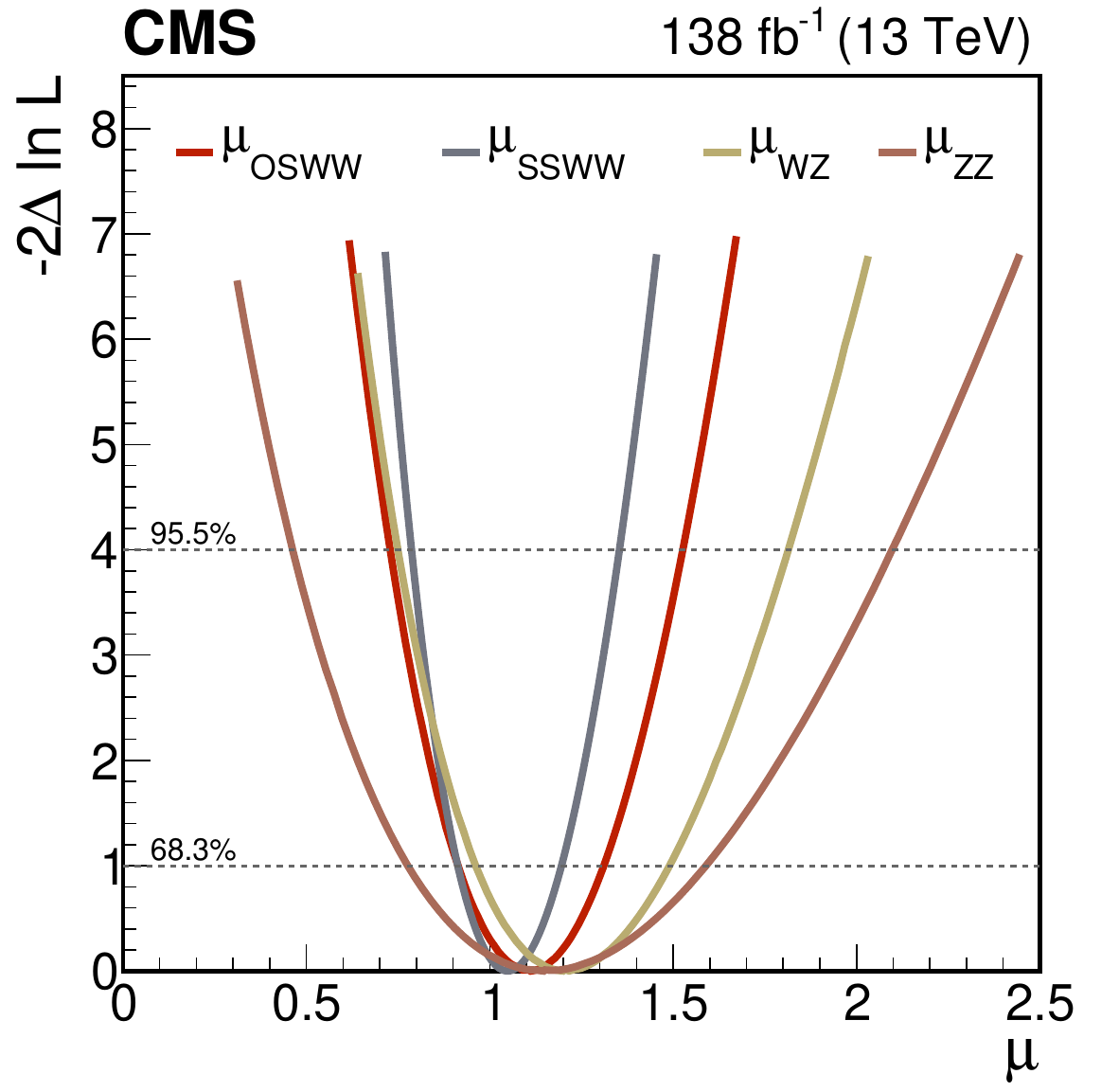}
    \includegraphics[width=0.45\textwidth]{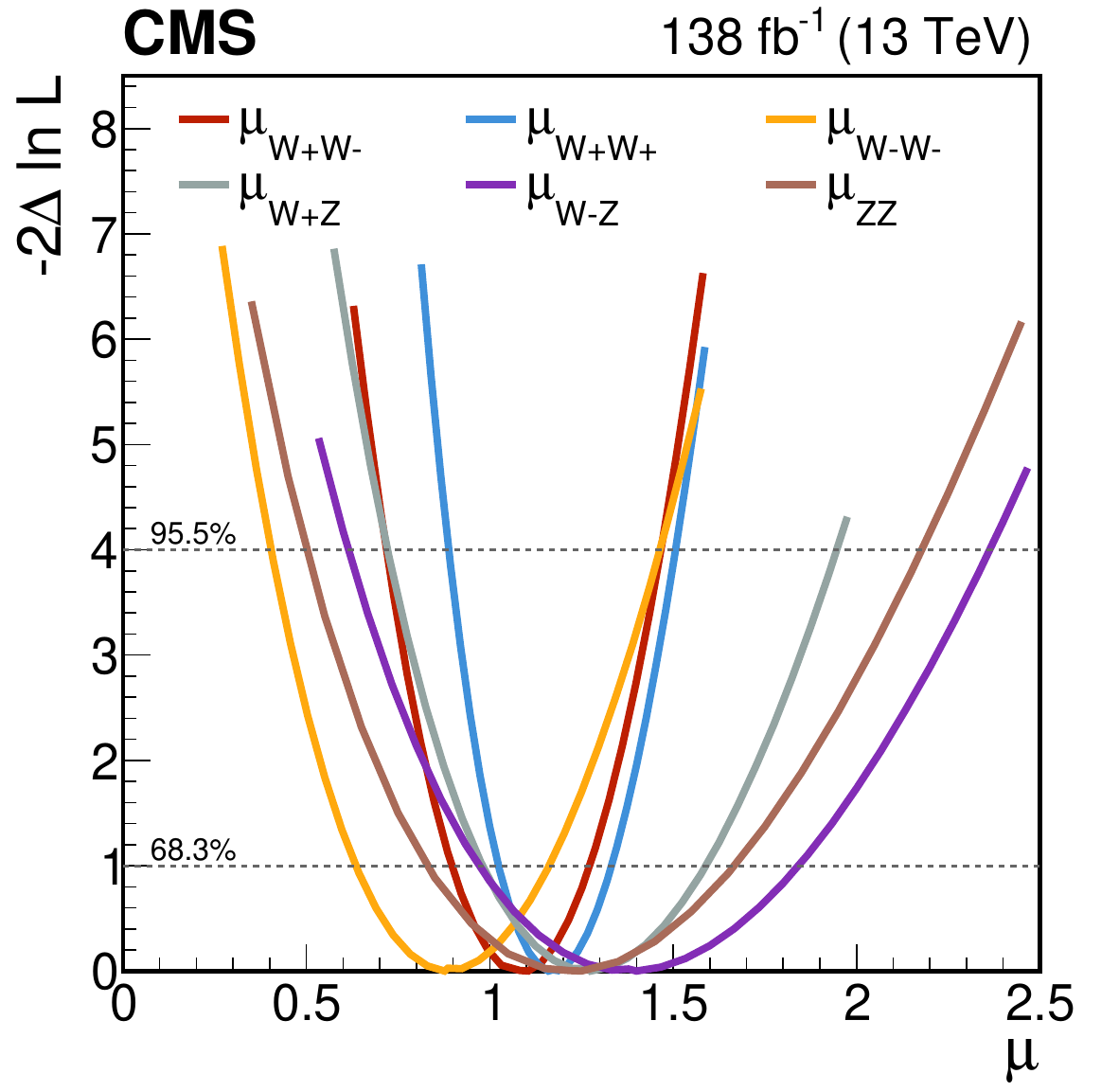}
    \caption{The likelihood profiles from the combined fit for the various POIs listed in the legends vs. $\mu = \sigma_{\text{obs}}/\sigma_{\text{SM}}$. The 68.3 ans 95.5\% \CL{}s are drawn as dashed horizontal lines and are computed as $-2\Delta \ln L = 1$ and $-2\Delta \ln L = 4$ respectively. The left (right) figure shows the $-2\Delta \ln L$ profiles for the 4-POIs (6-POIs) model.}\label{fig:likelihood_profiles_1D}
 \end{figure*}

Results are presented separately for the maximum likelihood fits performed considering two statistical models with four or six freely varying parameters. Table~\ref{tab:signal_strengths} summarizes the parameter estimates from the combined profile-likelihood fit, along with the observed and expected 68.3\% \CL intervals. The quoted statistical significance is defined, within the asymptotic approximation~\cite{Cowan:2010js}, as the probability of observing a larger deviation from the background-only hypothesis. The likelihood profiles are depicted in Fig.~\ref{fig:likelihood_profiles_1D}.
The improvement in the signal strength uncertainty is channel-dependent, reaching up to 15\% for the OS$\PW\PW$ and $\PW\PZ$ channels with respect to the best individual CMS measurements entering the combination~\cite{CMS:2022woe,CMS:2020gfh}.

\begin{table*}[!htb]
   \centering
   \topcaption{\label{tab:signal_strengths} The signal strength $\mu = \sigma_{\text{obs}}/\sigma_{\text{SM}}$, where $\sigma_{\text{obs}}$ and $\sigma_{\text{SM}}$ are the observed and SM predicted cross sections respectively, is reported for each VBS production mode (columns). Statistical significances $\sigma$ (in standard deviations) are also reported. The upper block reports the results for the model with four signal strengths while the lower block reports the same results but splitting the signal strengths for the \PW boson electric charge. The expected results are reported within parentheses. For the expected signal strengths, only the ${\pm}1$ standard deviation uncertainty is reported as the central value is assumed to be 1.}
   {\fontsize{8.5}{10}\selectfont
   \renewcommand{\arraystretch}{2.0}
   \setlength{\tabcolsep}{4pt}
   \begin{tabular}{ p{0.3cm}  c  c  c  c  c  c  c }
   \hline
   \parbox[t]{3mm}{\multirow{3}{*}{\rotatebox[origin=c]{90}{4-POIs}}} & & \multicolumn{2}{c}{SS$\PW\PW$} & OS$\PW\PW$  & \multicolumn{2}{c}{$\PW\PZ$}  & $\PZ\PZ$ \\

   \cline{2-8}

    & $\mu$ & \multicolumn{2}{c}{$1.05_{-0.14}^{+0.15}$ $\left({}_{-0.14}^{+0.14}\right)$} & $1.11_{-0.20}^{+0.20}$ $\left({}_{-0.18}^{+0.19}\right)$ & \multicolumn{2}{c}{$1.21_{-0.25}^{+0.28}$ $\left({}_{-0.24}^{+0.27}\right)$} & $1.15_{-0.37}^{+0.44}$ $\left({}_{-0.37}^{+0.44}\right)$\\

    & $\sigma$ & \multicolumn{2}{c}{$\gg$5 ($\gg$5)} & 6.4 (6.1) & \multicolumn{2}{c}{7.0 (5.6)} & 4.0 (3.6) \\

   \hline

   \parbox[t]{2mm}{\multirow{3}{*}{\rotatebox[origin=c]{90}{6-POIs}}} & & $\PWp\PWp$ & $\PWm\PWm$ & $\PWp\PWm$ & $\PWp\PZ$ & $\PWm\PZ$ & $\PZ\PZ$ \\
   \cline{2-8}
    & $\mu$ & $ 1.13_{-0.16}^{+0.17}$ $\left({}_{-0.14}^{+0.16}\right)$ & $0.84_{-0.24}^{+0.28}$ $\left({}_{-0.25}^{+0.28}\right)$ & $1.10_{-0.20}^{+0.20}$ $\left({}_{-0.18}^{+0.18}\right)$ & $1.16_{-0.27}^{+0.32}$ $\left({}_{-0.27}^{+0.32}\right)$ & $1.31_{-0.40}^{+0.47}$ $\left({}_{-0.37}^{+0.44}\right)$ & $1.16_{-0.38}^{+0.44}$ $\left({}_{-0.35}^{+0.42}\right)$ \\

    & $\sigma$ & $\gg$5 ($\gg$5) & 3.9 (4.6) & 6.3 (6.1)  & 5.7 (4.7)  & 4.2 (3.2) & 4.0 (3.6) \\
   \hline
   \end{tabular}
   }
\end{table*}

\begin{figure*}[!htb]
   \centering
   \includegraphics[width=0.70\textwidth]{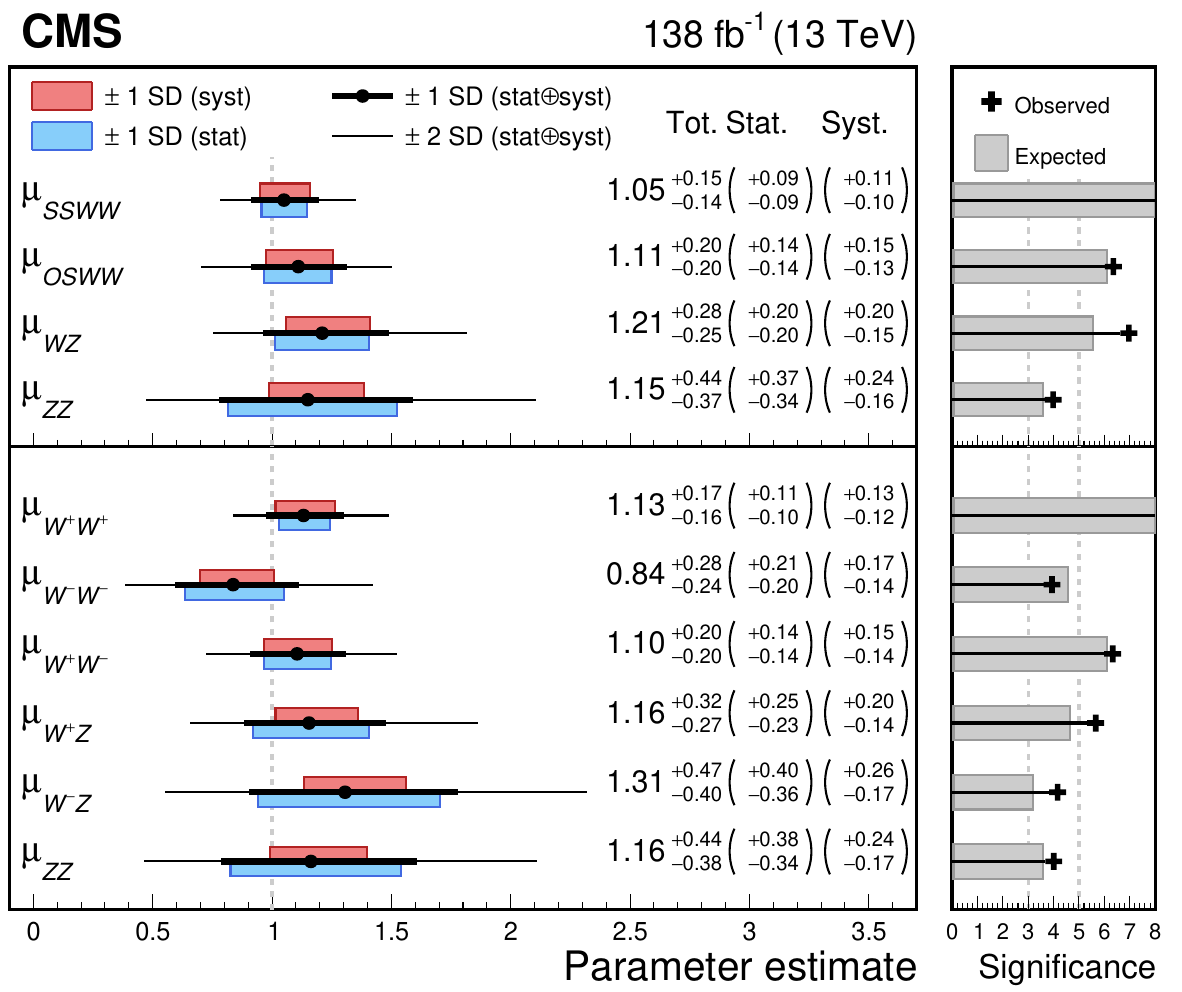}
   \caption{Measurement of the VBS signal strengths ($\mu$) and comparison with the SM predictions. The upper (lower) panel presents results considering four (six) independent signal strengths. In each panel, the left side shows the measured signal strengths, with thick (thin) black lines representing the 1 (2) standard deviation (SD) confidence intervals. The red and blue bands within the 1 standard deviation interval illustrate the systematic and statistical uncertainties, respectively. The right side displays the observed (black crosses) and expected (filled grey bars) statistical significance of the EW VBS signal relative to the SM background. The SS$\PW\PW$ and $\PWp\PWp$ significance bars are truncated, as their values exceed the axis range.} \label{fig:final_summary_1D}
\end{figure*}

The same information is reported graphically in Fig.~\ref{fig:final_summary_1D} with an additional breakdown for the statistical and systematic components of the uncertainties.  The measured signal strengths are in agreement with the SM predictions within the 68.3\% \CL.

\begin{figure*}[!htbp]
    \centering
    \includegraphics[width=0.45\textwidth]{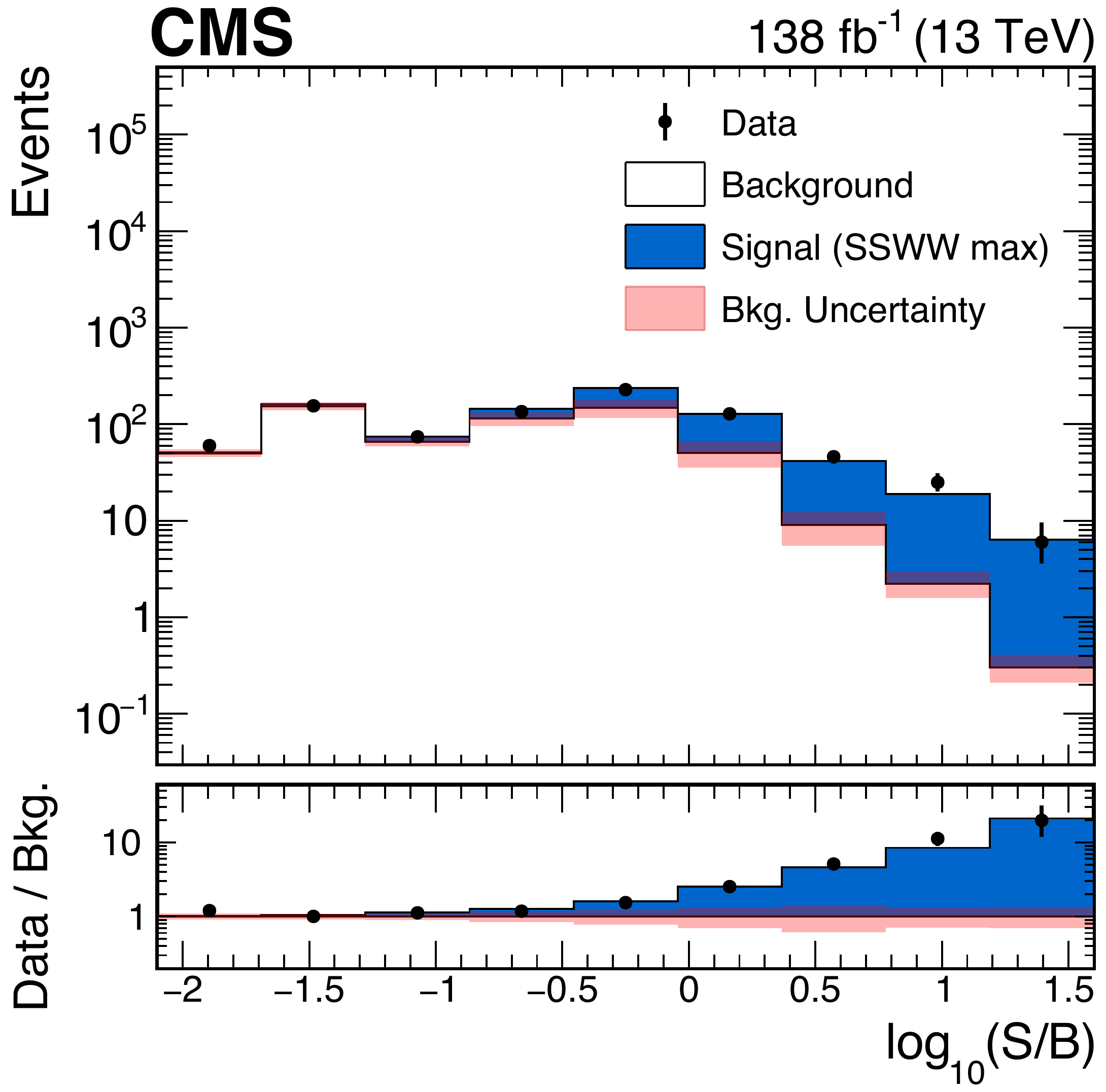}
    \includegraphics[width=0.45\textwidth]{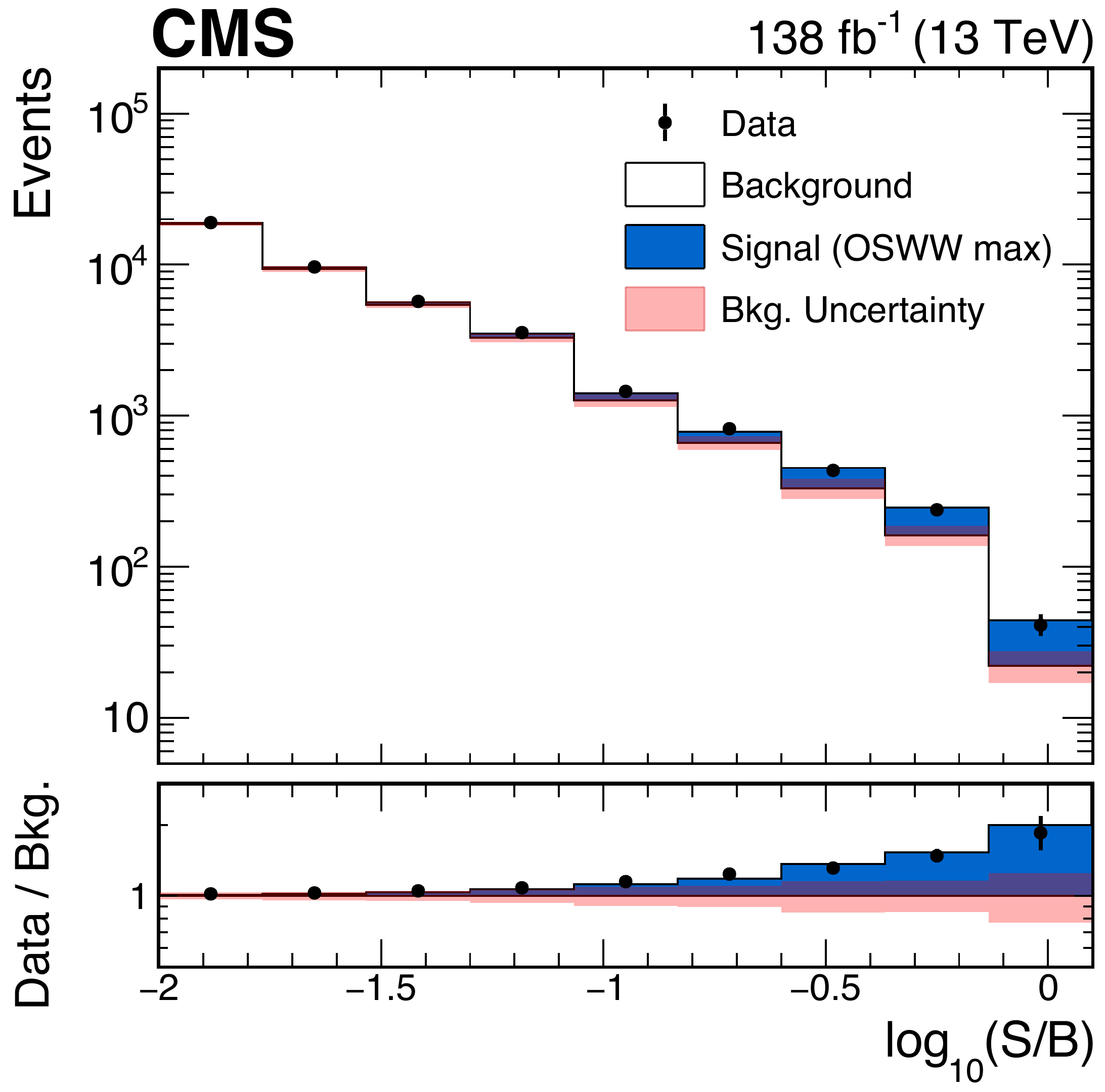}
    \includegraphics[width=0.45\textwidth]{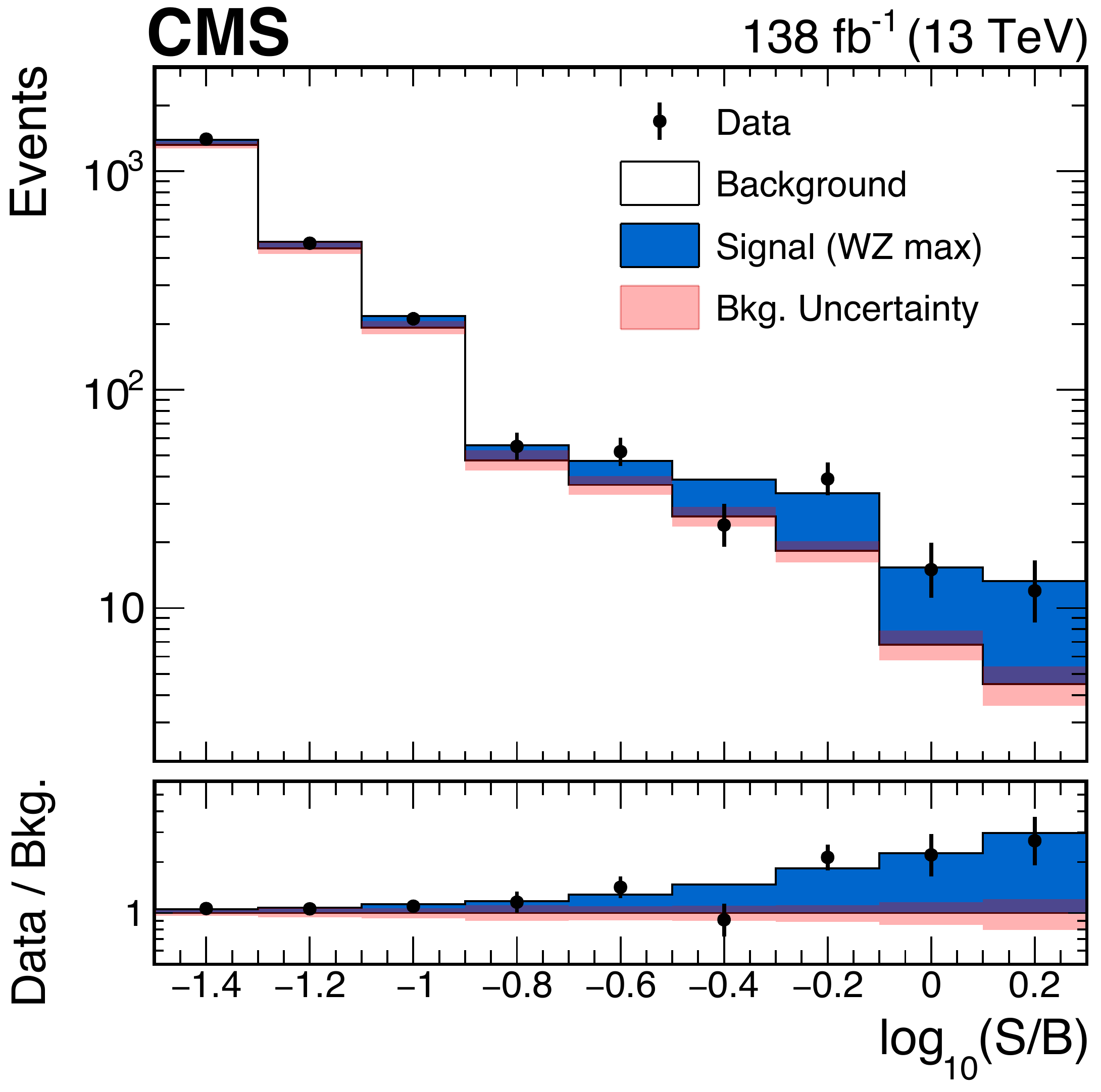}
    \includegraphics[width=0.45\textwidth]{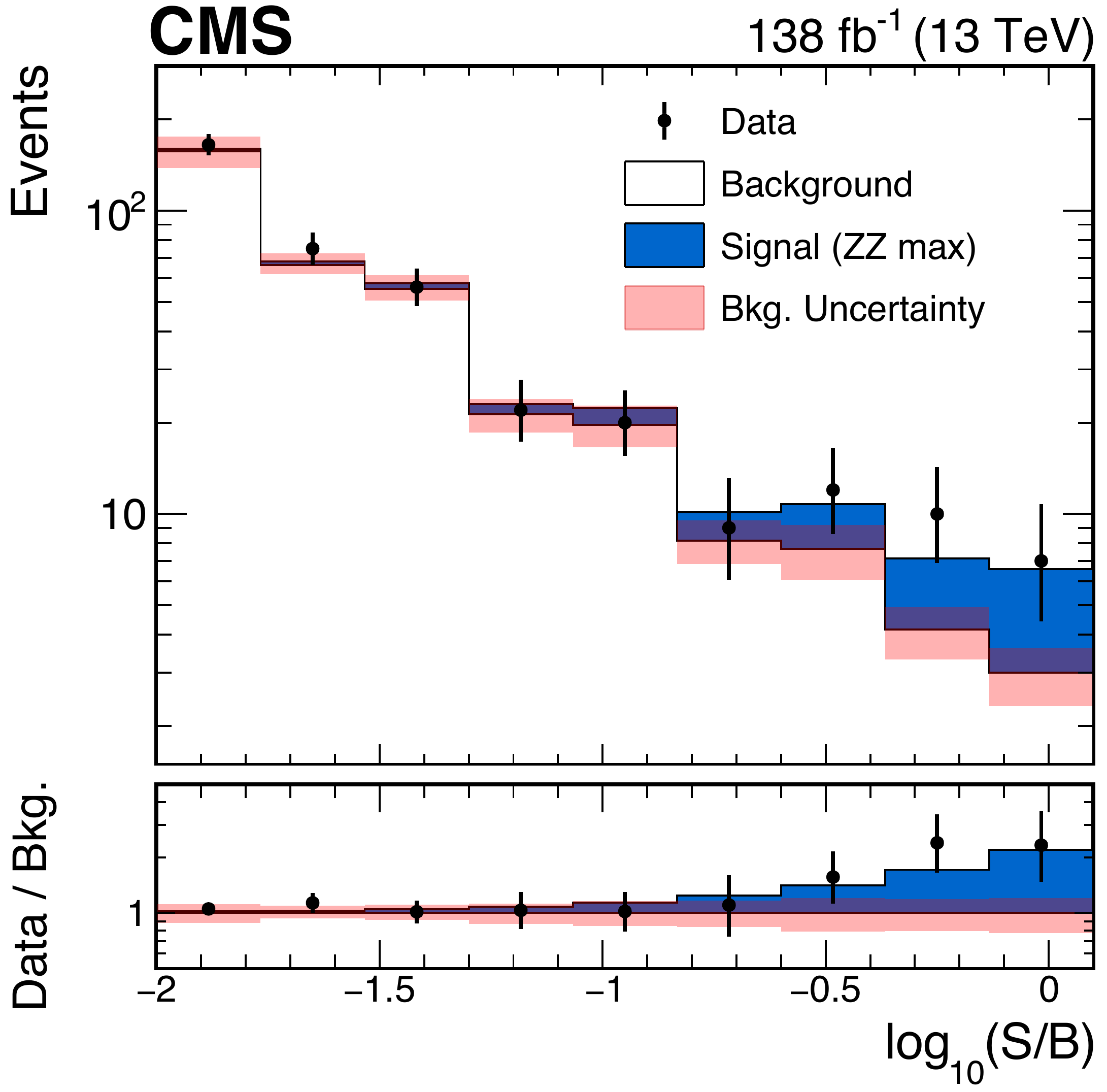}
    \caption{ Observed $\log_{10}[S(\mu_i=1)/B]$ distributions for the VBS SM combination and four parameters model. From left to right, upper to lower, the filled white histograms represent the overall background contributions, whereas the blue histograms represent the signal yields for $\mu_{\mathrm{SS}\PW\PW}$, $\mu_{\mathrm{OS}\PW\PW}$, $\mu_{\PW\PZ}$, and $\mu_{\PZ\PZ}$. Observations are displayed as black markers with vertical bars representing the uncertainty. The post-fit 68.3\% uncertainty interval on the overall background template is shown in pink. The distribution is binned as a function of the pre-fit value of $\log_{10}[S(\mu_i=1)/B]$, obtained from the input template bins, and is filled with the post-fit yields of total signal and backgrounds. Each input bin is assigned to the signal process providing the largest (``max") contribution in that bin of the input templates, as indicated in the legend. }\label{fig:s_o_b_plots}
\end{figure*}

To better visualize the result of the VBS combination in terms of signal excess over the SM background, distributions of the signal-to-background ratio ($S/B$) are constructed and shown in Fig.~\ref{fig:s_o_b_plots}. Four separate $S/B$ distributions are defined, corresponding to the four signal processes considered in the 4-POIs model. For each bin $k$ of the input templates, the quantity $\log_{10}\left[S(\vec{\mu}=1)/B\right]_k$ is computed prior to the combined fit, where $S$ denotes the total signal contribution. Each bin $k$ is associated with the $S/B$ distribution corresponding to the signal process that provides the largest contribution in that bin. The post-fit values of the background, total signal, and data yields are then filled into the corresponding $\log_{10}(S/B)_k$ bin of the selected distribution. The uncertainties in the bins of the $S/B$ distributions are estimated using pseudo-experiments, from which the central 68.3\% interval is extracted.

\begin{figure*}[!htbp]
    \centering
    \includegraphics[width=0.45\textwidth]{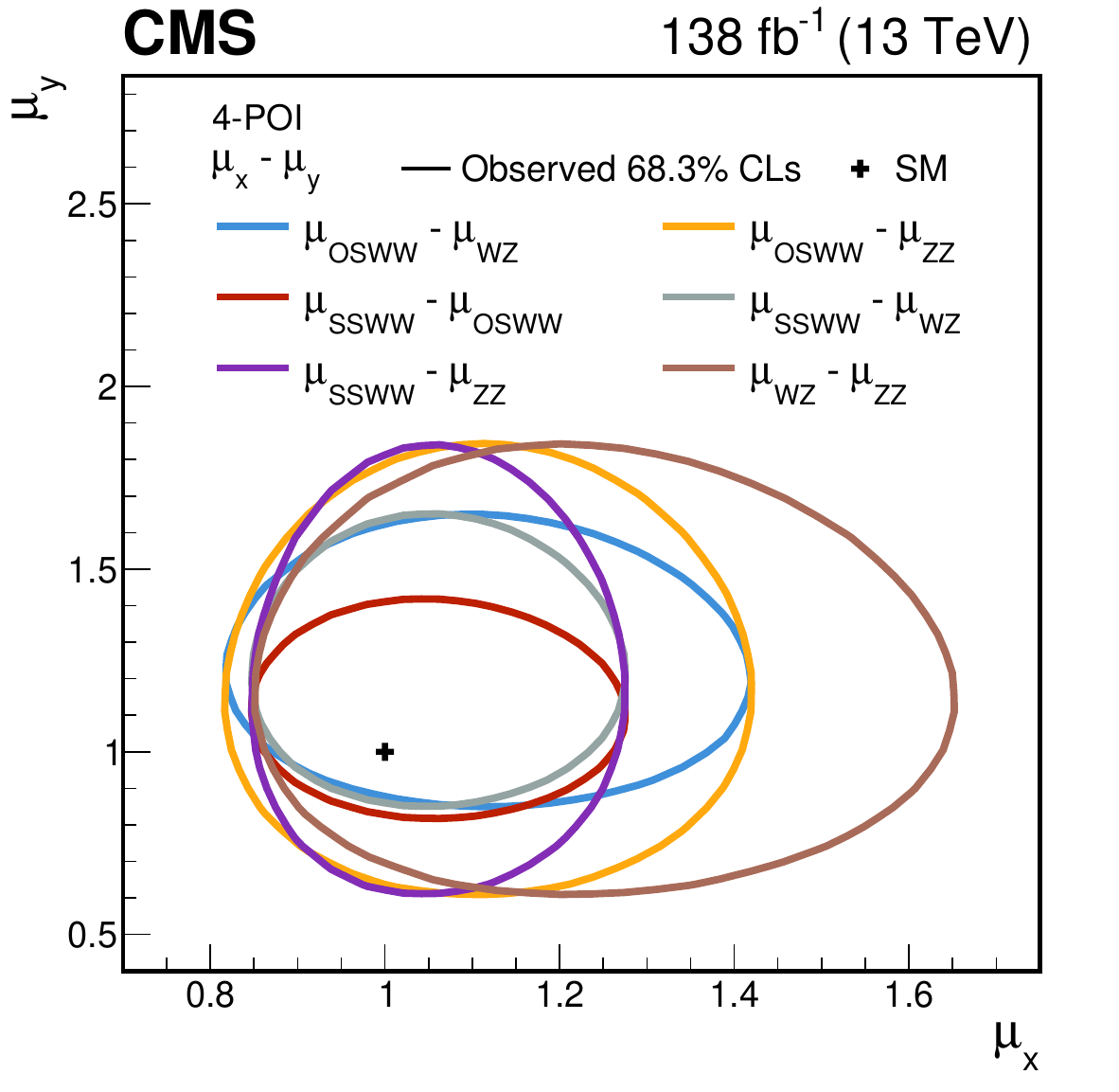}
    \includegraphics[width=0.45\textwidth]{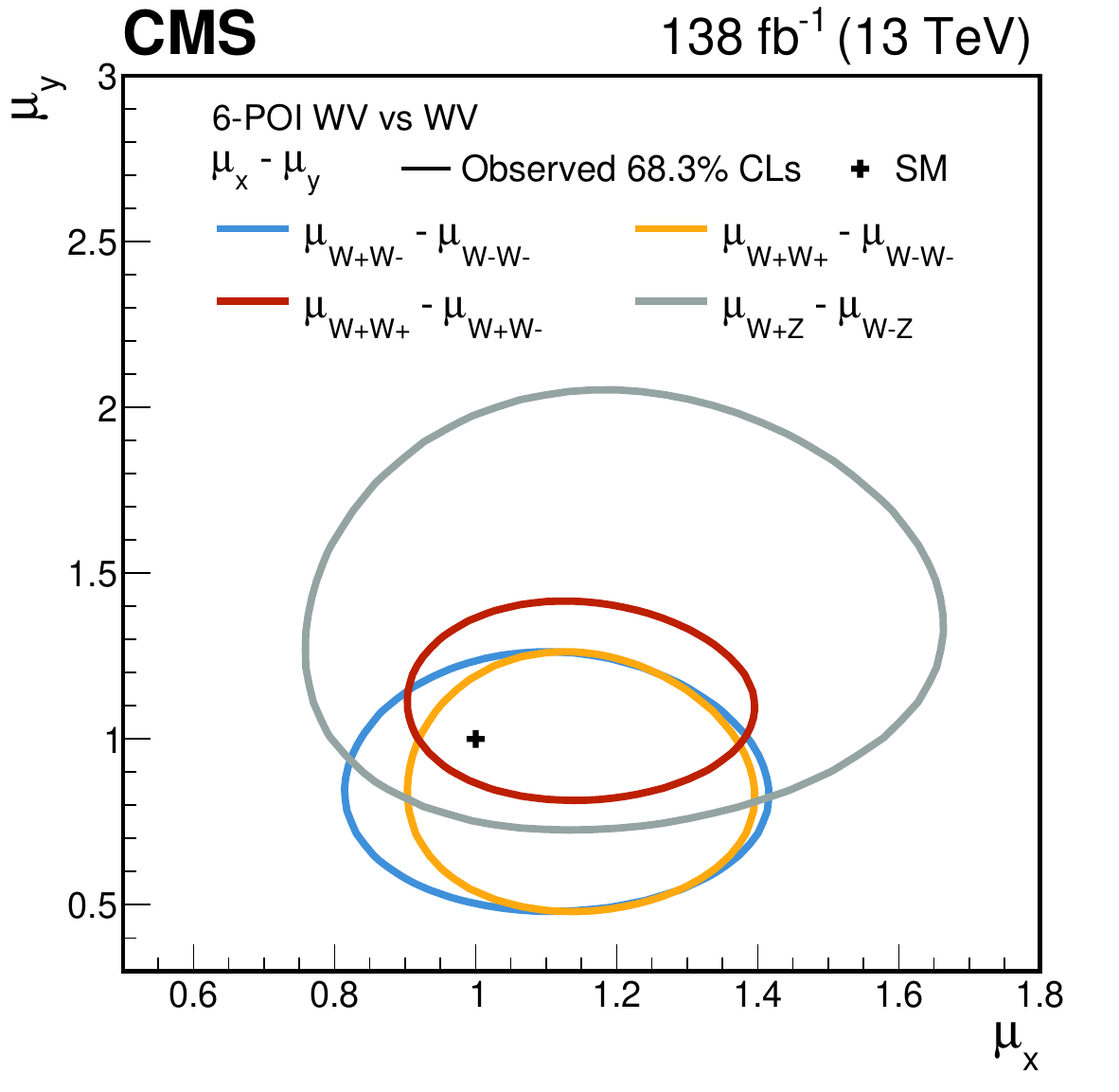}
    \includegraphics[width=0.45\textwidth]{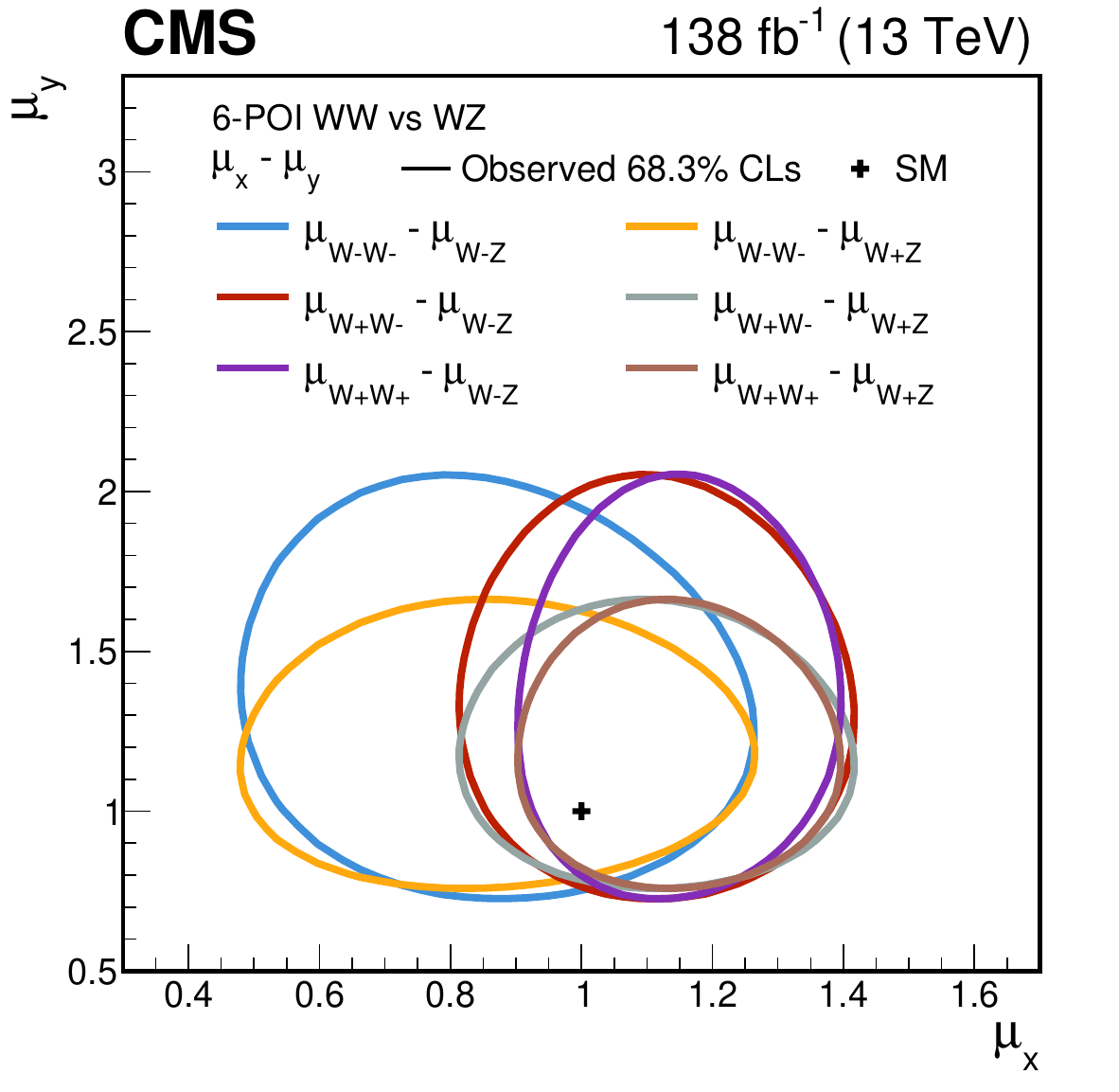}
    \includegraphics[width=0.45\textwidth]{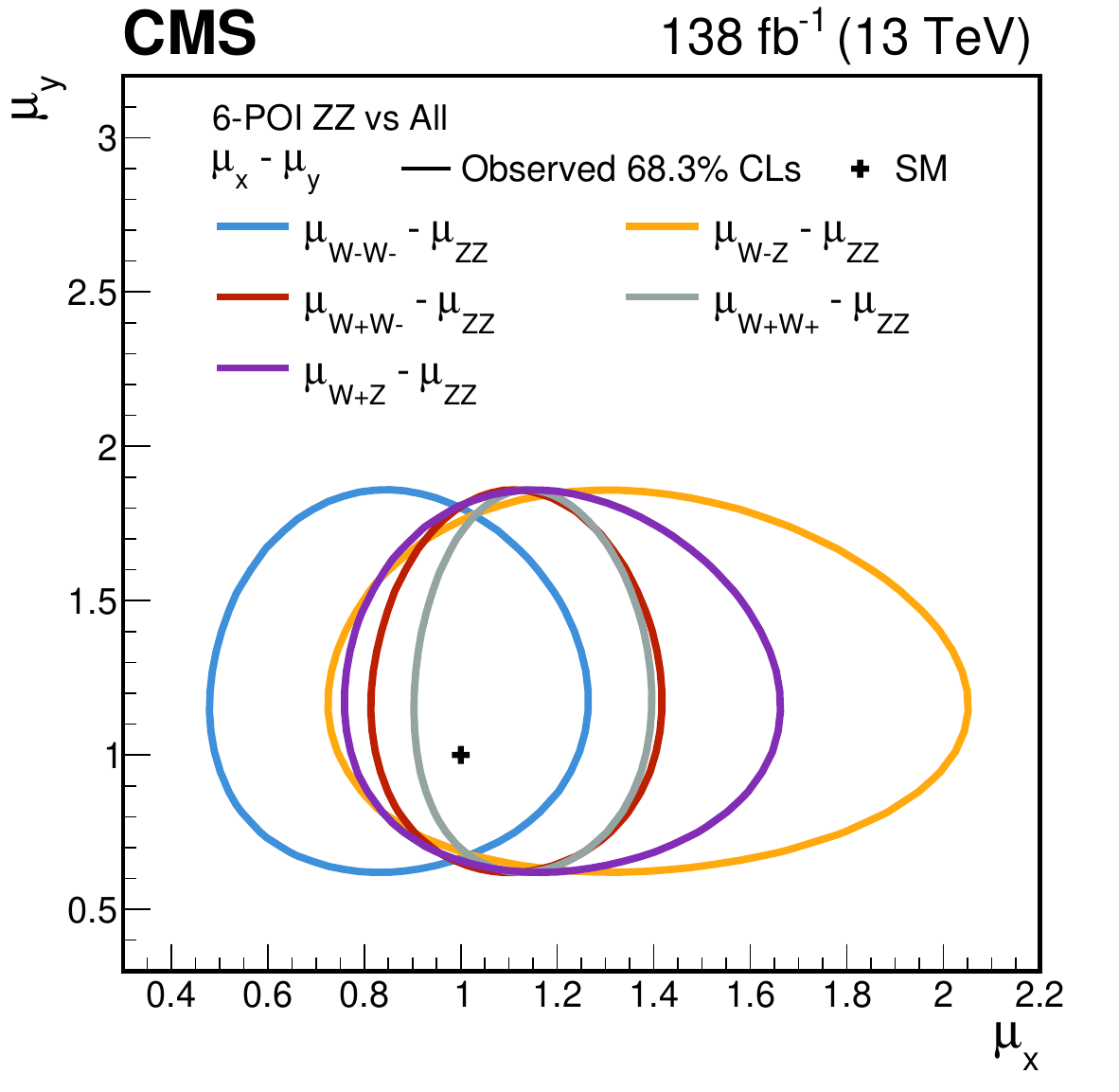}
    \caption{Summary plots of the two-dimensional fits for all pairs of POIs $\mu = \sigma_{\text{obs}}/\sigma_{\text{SM}}$. The 68.3\% \CL{}s, drawn as solid lines, are computed from the likelihood profiles as $-2\Delta \ln L < 2.3$. The upper-left panel shows the 68.3\% \CL intervals for all pairs of parameters of interest in the 4-POIs model, with the legend listing first the parameter on the x axis, followed by that on the y axis. The results for the 6-POIs model are split across three panels: the upper-right panel shows all pairs involving $\PW\PV$ combinations ($\PW\PW$ and $\PW\PZ$), the lower-left panel shows all combinations of $\PW\PW$ and $\PW\PZ$ parameters, and the lower-right panel shows all pairs involving $\PZ\PZ$. In all cases, all parameters of interest other than the two under consideration are profiled in the maximum likelihood fit, together with the nuisance parameters.} \label{fig:likelihood_profiles_2D}
 \end{figure*}

Simultaneous fits to all pairs of signal strengths are performed to assess the correlation between the different VBS production modes. As for the single POI case, all other POIs except the ones of interest are profiled in the maximum likelihood fit along with the nuisance parameters. The two-dimensional results for both 4 and 6 parameters are in agreement with the SM expectation and present mild correlations among all signal strengths, as shown in Fig.~\ref{fig:likelihood_profiles_2D}.

\section{Summary}

The most comprehensive statistical combination of vector boson scattering (VBS) processes to date, integrating results from seven distinct analyses using CMS data, has been presented. Unlike previous single-channel measurements, this global combination accounts for the correlations between all VBS processes involving massive vector bosons ($\PW$, $\PZ$), providing a unique and rigorous stress test of the standard model (SM) electroweak (EW) sector. The targeted VBS production modes are all charge combinations of $\PW\PW$, $\PW\PZ$, and $\PZ\PZ$, with $\PW\PW$ further split into same-sign (SSWW) and opposite-sign (OSWW) channels. Each process is studied in final states where both vector bosons decay leptonically into electrons, muons, or tau leptons, as well as in final states where one vector boson decays hadronically.

The EW signal strengths are measured via a simultaneous fit of all channels in two separate configurations: one merging W boson electric charges with four free parameters ($\mu_{\mathrm{SS}\PW\PW}$, $\mu_{\mathrm{OS}\PW\PW}$, $\mu_{\PW\PZ}$, $\mu_{\PZ\PZ}$) and another separating \PW boson charges with six parameters ($\mu_{\PWp\PWp}$, $\mu_{\PWm\PWm}$, $\mu_{\PWp\PWm}$, $\mu_{\PWp\PZ}$, $\mu_{\PWm\PZ}$, $\mu_{\PZ\PZ}$).

Simultaneous one- and two-dimensional fits were performed for all VBS EW production components. Overall, all results are found to be consistent with the SM predictions. 

This combination provides a simultaneous constraint of all VBS processes involving massive vector bosons, including charge-separated channels, enabling a coherent test of the EW gauge structure in VBS that was not accessible in previous single-channel measurements. It improves the precision of VBS measurements and provides a benchmark for future VBS studies at the LHC.

\ifthenelse{\boolean{cms@external}}{\clearpage}{}
\begin{acknowledgments}
We congratulate our colleagues in the CERN accelerator departments for the excellent performance of the LHC and thank the technical and administrative staffs at CERN and at other CMS institutes for their contributions to the success of the CMS effort. In addition, we gratefully acknowledge the computing centers and personnel of the Worldwide LHC Computing Grid and other centers for delivering so effectively the computing infrastructure essential to our analyses. Finally, we acknowledge the enduring support for the construction and operation of the LHC, the CMS detector, and the supporting computing infrastructure provided by the following funding agencies: SC (Armenia), BMFWF and FWF (Austria); FNRS and FWO (Belgium); CNPq, CAPES, FAPERJ, FAPERGS, and FAPESP (Brazil); MES and BNSF (Bulgaria); CERN; CAS, MoST, and NSFC (China); MINCIENCIAS (Colombia); MSES and CSF (Croatia); RIF (Cyprus); SENESCYT (Ecuador); ERC PRG and PSG, TARISTU24-TK10 and MoER TK202 (Estonia); Academy of Finland, MEC, and HIP (Finland); CEA and CNRS/IN2P3 (France); SRNSF (Georgia); BMFTR, DFG, and HGF (Germany); GSRI (Greece); MATE and NKFIH (Hungary); DAE and DST (India); IPM (Iran); SFI (Ireland); INFN (Italy); MSIT and NRF (Republic of Korea); MES (Latvia); LMTLT (Lithuania); MOE and UM (Malaysia); BUAP, CINVESTAV, CONACYT, LNS, SEP, and UASLP-FAI (Mexico); MOS (Montenegro); MBIE (New Zealand); PAEC (Pakistan); MSHE, NSC, and NAWA (Poland); FCT (Portugal); MESTD (Serbia); MICIU/AEI and PCTI (Spain); MOSTR (Sri Lanka); Swiss Funding Agencies (Switzerland); MST (Taipei); MHESI (Thailand); TUBITAK and TENMAK (T\"{u}rkiye); NASU (Ukraine); STFC (United Kingdom); DOE and NSF (USA).

\begin{sloppypar}
\setlength\emergencystretch{\hsize}
\hyphenation{Rachada-pisek} Individuals have received support from the Marie-Curie program and the European Research Council and Horizon 2020 Grant, contract Nos.\ 675440, 724704, 752730, 758316, 765710, 824093, 101115353, 101002207, 101001205, and COST Action CA16108 (European Union); the Leventis Foundation; the Alfred P.\ Sloan Foundation; the Alexander von Humboldt Foundation; the Science Committee, project no. 22rl-037 (Armenia); the Fonds pour la Formation \`a la Recherche dans l'Industrie et dans l'Agriculture (FRIA) and Fonds voor Wetenschappelijk Onderzoek contract No. 1228724N (Belgium); the Beijing Municipal Science \& Technology Commission, No. Z191100007219010, the Fundamental Research Funds for the Central Universities, the Ministry of Science and Technology of China under Grant No. 2023YFA1605804, the Natural Science Foundation of China under Grant No. 12535004, and USTC Research Funds of the Double First-Class Initiative No.\ YD2030002017 (China); the Ministry of Education, Youth and Sports (MEYS) of the Czech Republic; the Shota Rustaveli National Science Foundation (Georgia); the Deutsche Forschungsgemeinschaft (DFG), among others, under Germany's Excellence Strategy -- EXC 2121 ``Quantum Universe" -- 390833306, and under project number 400140256 - GRK2497; the Hellenic Foundation for Research and Innovation (HFRI), Project Number 2288 (Greece); the Hungarian Academy of Sciences, the New National Excellence Program - \'UNKP, the NKFIH research grants K 131991, K 138136, K 143460, K 143477, K 147557, K 146913, K 146914, K 147048, TKP2021-NKTA-64, and 2025-1.1.5-NEMZ\_KI-2025-00004, and MATE KKP and KKPCs Research Excellence and Flagship Research Groups grants (Hungary); the Council of Science and Industrial Research, India; ICSC -- National Research Center for High Performance Computing, Big Data and Quantum Computing, FAIR -- Future Artificial Intelligence Research, and CUP I53D23001070006 (Mission 4 Component 1), funded by the NextGenerationEU program, the Italian Ministry of University and Research (MUR) under Bando PRIN 2022 -- CUP I53C24002390006, PRIN PRIMULA 2022RBYK7T (Italy); the Latvian Council of Science; the Ministry of Science and Higher Education, project no. 2022/WK/14, and the National Science Center, contracts Opus 2021/41/B/ST2/01369, 2021/43/B/ST2/01552, 2023/49/B/ST2/03273, and the NAWA contract BPN/PPO/2021/1/00011 (Poland); the Funda\c{c}\~ao para a Ci\^encia e a Tecnologia (Portugal); the National Priorities Research Program by Qatar National Research Fund; MICIU/AEI/10.13039/501100011033, ERDF/EU, ``European Union NextGenerationEU/PRTR", projects PID2022-142604OB-C21, PID2022-139519OB-C21, PID2023-147706NB-I00, PID2023-148896NB-I00, PID2023-146983NB-I00, PID2023-147115NB-I00, PID2023-148418NB-C41, PID2023-148418NB-C42, PID2023-148418NB-C43, PID2023-148418NB-C44, PID2024-158190NB-C22, RYC2021-033305-I, RYC2024-048719-I, CNS2023-144781, CNS2024-154769 and Plan de Ciencia, Tecnolog{\'i}a e Innovaci{\'o}n de Asturias, Spain; the Chulalongkorn Academic into Its 2nd Century Project Advancement Project, the National Science, Research and Innovation Fund program IND\_FF\_68\_369\_2300\_097, and the Program Management Unit for Human Resources \& Institutional Development, Research and Innovation, grant B39G680009 (Thailand); the Eric \& Wendy Schmidt Fund for Strategic Innovation through the CERN Next Generation Triggers project under grant agreement number SIF-2023-004; the Kavli Foundation; the Nvidia Corporation; the SuperMicro Corporation; the Welch Foundation, contract C-1845; and the Weston Havens Foundation (USA).
\end{sloppypar}
\end{acknowledgments}\section*{Data availability} Release and preservation of data used by the CMS Collaboration as the basis for publications is guided by the  \href{https://doi.org/10.7483/OPENDATA.CMS.1BNU.8V1W}{CMS data preservation, re-use and open access policy}.
\bibliography{auto_generated}

\cleardoublepage \appendix\section{The CMS Collaboration \label{app:collab}}\begin{sloppypar}\hyphenpenalty=5000\widowpenalty=500\clubpenalty=5000\cmsinstitute{Yerevan Physics Institute, Yerevan, Armenia}
{\tolerance=6000
A.~Gevorgyan\cmsorcid{0000-0003-2751-9489}, A.~Hayrapetyan, V.~Makarenko\cmsorcid{0000-0002-8406-8605}, A.~Tumasyan\cmsAuthorMark{1}\cmsorcid{0009-0000-0684-6742}
\par}
\cmsinstitute{Institut f\"{u}r Hochenergiephysik, Vienna, Austria}
{\tolerance=6000
P.S.~Hussain\cmsorcid{0000-0002-4825-5278}, M.~Sonawane\cmsorcid{0000-0003-0510-7010}
\par}
\cmsinstitute{Marietta Blau Institute for Particle Physics, Vienna, Austria}
{\tolerance=6000
W.~Adam\cmsorcid{0000-0001-9099-4341}, L.~Benato\cmsorcid{0000-0001-5135-7489}, T.~Bergauer\cmsorcid{0000-0002-5786-0293}, M.~Dragicevic\cmsorcid{0000-0003-1967-6783}, S.~Gundacker\cmsorcid{0000-0003-2087-3266}, A.K.~Guven\cmsorcid{0009-0004-5670-5138}, M.~Jeitler\cmsAuthorMark{2}\cmsorcid{0000-0002-5141-9560}, N.~Krammer\cmsorcid{0000-0002-0548-0985}, A.~Li\cmsorcid{0000-0002-4547-116X}, D.~Liko\cmsorcid{0000-0002-3380-473X}, M.~Matthewman, A.~Pfeiffer\cmsorcid{0000-0001-5328-448X}, J.~Schieck\cmsAuthorMark{2}\cmsorcid{0000-0002-1058-8093}, R.~Sch\"{o}fbeck\cmsAuthorMark{2}\cmsorcid{0000-0002-2332-8784}, M.~Shooshtari\cmsorcid{0009-0004-8882-4887}, N.~Van~Den~Bossche\cmsorcid{0000-0003-2973-4991}, W.~Waltenberger\cmsorcid{0000-0002-6215-7228}, C.E.~Wulz\cmsAuthorMark{2}\cmsorcid{0000-0001-9226-5812}
% ...
\par}
$^{1}$Also at Yerevan State University, Yerevan, Armenia\\
$^{2}$Also at Technische Universit\"{a}t Wien, Vienna, Austria\\
% ...
\end{sloppypar}
\end{document}